\documentclass[aps,pre,twocolumn,groupedaddress]{revtex4}

\usepackage{graphicx}

\begin{document}

\def\hslash{\hbar}
\def\imag{i}
\def\grad{\vec{\nabla}}
\def\div{\vec{\nabla}\cdot}
\def\curl{\vec{\nabla}\times}
\def\DDt{\frac{d}{dt}}
\def\ddt{\frac{\partial}{\partial t}}
\def\ddx{\frac{\partial}{\partial x}}
\def\ddy{\frac{\partial}{\partial y}}
\def\lap{\nabla^{2}}
\def\divv{\vec{\nabla}\cdot\vec{v}}
\def\gradS{\vec{\nabla}S}
\def\vvec{\vec{v}}
\def\wc{\omega_{c}}
\def\<{\langle}
\def\>{\rangle}
\def\Tr{{\rm Tr}}
\def\Csch{{\rm csch}}
\def\Coth{{\rm coth}}
\def\Tanh{{\rm tanh}}
\def\g2{g^{(2)}}

\title{Exciton and charge-transfer dynamics in polymer semiconductors}

\author{Eric R. Bittner}
\affiliation{Department of Chemistry and Center for Materials Chemisty, University of Houston, Houston, TX}

\author{John Glen S. Ramon}
\affiliation{Department of Chemistry and Center for Materials Chemisty, University of Houston, Houston, TX}

\date{\today}

\begin{abstract}
Organic semiconducting polymers are currently of broad interest as
potential low-cost materials for photovoltaic and light-emitting
display applications. I will give an overview of our work in
developing a consistent quantum dynamical picture of the excited state
dynamics underlying the photo-physics. We will also focus upon the
quantum relaxation and reogranization dynamics that occur upon
photoexcitation of a couple of type II donor-acceptor polymer
heterojunction systems. Our results stress the significance of
vibrational relaxation in the state-to- state relaxation and the
impact of curve crossing between charge- transfer and excitonic
states. Furthermore, while a tightly bound charge-transfer state
(exciplex) remain the lowest excited state, we show that the
regeneration of the optically active lowest excitonic state in
TFB:F8BT is possible via the existence of a steady-state involving the
bulk charge-transfer state. Finally, we will discuss ramifications of
these results to recent experimental studied and the fabrication of
efficient polymer LED and photovoltaics.\end{abstract}

\maketitle              % typeset the title of the contribution

\section{Introduction}

\begin{figure}
\includegraphics[width=\columnwidth]{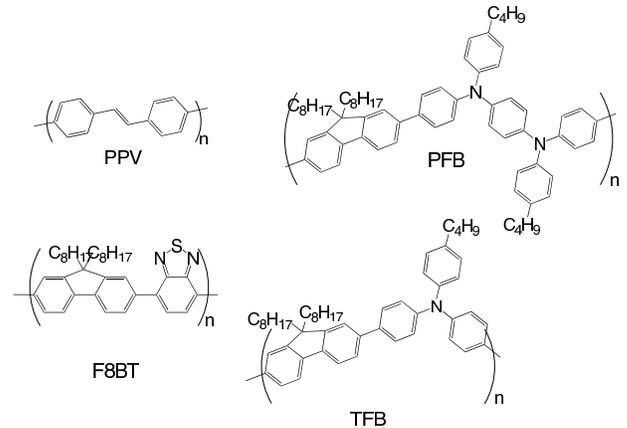}
\caption{Structures and common short-hand names of various conjugated poly-phenylene
derived semiconducting polymers that are of interest for fabricating luminescent devices.}\label{polymers}
\end{figure}

Organic semiconducting materials have attracted the attention of the
electronic industry.  While the low charge mobility compared to
crystalline silicon and various other characteristics limit the range
of potential application for organics, a wide range of properties such
as solubility in organic solvents and the color of light emission can
be finely tuned using ordinary chemical synthesis.  Molecules and
their arrangement can be easily engineered to fit a specific
requirement.  Moreover, polymer coatings can be easily applied to a
wide range of surfaces and media, including mechanically flexible
ones, to fabricate large scale displays or electronic sensors with
complex shapes.  Organic electronics complement and extend the range
of traditional silicon based electronics.  The chemical structures of
various common semiconducting polymer chains are given in
Fig~\ref{polymers}.  In practice, PPV is often functionalized with
alkyl side chains in order to tune its electronic properties and
increase its solubility in organic solvents.

Organic LED devices are typically layered structures with luminescent
media sandwiched between cathode and anode materials which are
selected such that their Fermi energies roughly match the conduction
and valence bands of the luminescent material.  Often the
semiconducting media itself consists of a hole transport layer and an
electron transport layer engineered to facilitate the rapid diffusion
of the injected carriers away from their image charges on the cathode
or anode.  These carriers are best described as polarons since
electron-phonon coupling produces significant lattice reorganization
about the carriers.  Finally, a third, luminescent layer can be
sandwiched between the transport layers.  In this layer, the electron
and hole polarons interact and combine to produce excitons.  The
individual spins of the electrons and holes are uncorrelated and only
singlet excitons are radiatively coupled the ground electronic state.
In the absence of singlet-triplet coupling, this places a theoretical
upper-limit or 1:4 or 25\% on the overall efficiency of a LED device
and it has been long debated whether or not the efficiency of organic
LED devices is in fact limited by this theoretical upper-limit.

The electronic properties of these materials are derived from the delocalized
$\pi$ orbitals found in conjugated polymers.  The $\pi$ electron
system is primarily an intramolecular network extending along the
polymer chain.  For a linear chain, the valance and conduction $\pi$
and $\pi^*$ bands are typically 1-3 eV wide compared to the
intermolecular bandwidth (due to $\pi$-stacking) of about 0.1eV for
well ordered materials.  Thus, intrachain charge transport is
extremely efficient; however, interchain transport typically limits
the charge mobility for the usual size range of devices.  The polymer
backbone is held together through a $\sigma$ bonding network.  These
bonds are considerably stronger than the $\pi$ bonds and keep the
molecule intact even following photoexcitation.  Hence, we can
consider the electronic dynamics as taking place within the $\pi$ band
and treat the localized $\sigma$ bonds as skeletal framework.

Since the dielectric constant of organic semiconductors is relatively low, 
screening between charges is relatively week.  At a
given radius, $r_c$, thermal fluctuation will be insufficient to break
apart an electron/hole pair,
$$
kT = \frac{e^2}{\epsilon r_c}
$$
at 300K, this radius is approximately 20nm, which is on the order of a
few molecular lengths.  If we consider the electron/hole pair to be a
hydrogenic-type system with effective masses equal to the free
electron mass and dielectric constant of 3, the resulting binding
energy is about 0.75eV with an effective Bohr radius of 0.3nm, which
effectively confines the exciton to a single molecular unit.  Finally,
if we consider the electron/hole pair to be pair of bound Fermions,
exchange energy resulting from the anti-symetrization of the
electron/hole wave function splits the spin-singlet and spin-triplet
excitons by about 0.5-0.7 eV with the spin-triplet lying lower in
energy than the singlet.  While both species are relatively localized,
singlets typically span about 10nm in well ordered materials while
triplets are much more localized.  In absence of spin-orbit coupling,
emission from the triplet states is forbidden.  Hence, triplet
formation in electron/hole capture can dramatically limit the
efficiency of a light-emitting diode device, although strong
theoretical and experimental evidence indicates that singlet formation
can be enhanced in long-chain polymers.

Experiments by various groups suggest that in long-chain conjugated
polymer systems, the singlet exciton population can be greatly
enhanced and that efficiencies as high as 60 to 80\% can be easily
achieved in PPV type systems.~\cite{friend:121} On the other hand, in
small oligomers, the theoretical upper limit appears to hold true.
These initial experiments were then followed by a remarkable set of
observations by Wohlgenannt and Vardeny~\cite{wohlgenannt:197401} that
indicate that the singlet to triplet ratio, $r > 1$ for wide range of
conjugated polymer systems and that $r$ scales universally with the
polaron energy--which itself scales inversely with the persistence
length of the $\pi$-conjugation,
\begin{eqnarray}
r \propto 1/n.
\end{eqnarray}
The electro-luminescent efficiency, $\phi$, is proportional to the actual singlet population
and is related to $r$ via
$$
\phi = r/(r+3). 
$$

Various mechanisms favoring the formation of singlets have been
proposed for both interchain and intrachain e-h collisions.  Using
Fermi's golden rule, Shuai, Bredas et
al.\cite{beljonne:419,ye:045208,shuai:131} indicate that the $S$ cross
section for interchain recombination can be higher than the triplet
one due to bond-charge correlations.  Wohlgenannt {\em et
al.}\cite{wohlgenannt:617} employ a similar model of two parallel
polyene chains.  Both of these works neglect vibronic and relaxation
effects.  In simulating the intrachain collision of opposite polarons,
Kobrak and Bittner \cite{kobrak:11473,kobrak:5399,kobrak:7684} show
that formation of singlets are enhanced by the near-resonance with the
free e-h pair.  The result reflects the fact that spin-exchange
renders the triplet more tightly bound than the singlet and hence more
electronic energy must be dissipated by the phonons in the formation
of the former. The energy-conservation constraints in spin-dependent
e-h recombination have been analyzed by Burin and
Ratner\cite{burin:6092} in an essential-state model. The authors point
out that nonradiative processes (internal conversion, intersystem
crossing) must entail C=C stretching vibrons since these modes couple
most strongly to $\pi\rightarrow\pi^*$ excitations.  Tandon et
al. suggest that irrespective of the recombination process, interchain
or intrachain, the {\em direct transition} to form singlets should
always be easier than triplets due to its smaller binding energy
relative to the triplet.~\cite{tandon:045109}.  A comprehensive review
of detailing the experiments and theory of this effect was presented
by Wohlgenannt and Vardeny,\cite{wohlgenannt:241201}.  By and large,
recent theoretical models point towards the role of multi-phonon
relaxation and the scaling of the singlet/triplet splitting with chain
length as dominant factors in determining this
enhancement.~\cite{wohlgenannt:165111,barford:205204,karabunarliev:3988,karabunarliev:057402}

If we assume that the electron/hole capture proceeds via a series of
microstates one can show that the ratio of the singlet to triplet
capture cross-sections, $r$ scale with the ratio of the exciton
binding energies\cite{bittner-ijqc}
$$
r\propto \frac{\epsilon_B^T}{\epsilon_B^S}.
$$
If we take the singlet-triplet energy difference to be equal to twice
the electron-hole exchange energy, $\Delta E_{ST} = 2 K$, and expand
$K$ in terms of the inverse conjugation length,
$$
K = K_\infty + K^{(1)}/n + \cdots
$$
where $2K_\infty$ is the singlet-triplet splitting of an infinitely
long polymer chain, one obtains
$$
r \propto 1 + n K_\infty/\epsilon_{B}^S + \cdots
$$
Since both $K_\infty > 0$, $r >1$.  Moreover, if we take $\epsilon_B^S
\propto 1/n$, we obtain a simple and universal scaling law for the
singlet-triplet capture ratio, $r \propto n$.  This ``universal''
scaling law for $r$ was reported by Wohlgennant and
Vardeny~\cite{wohlgenannt:197401}.  What is even more surprising, is
that the the same scaling law (i.e. slope and intercept for $r = a n +
c$) describes nearly all organic conjugated polymer systems.  Hence,
it appears that a common set of electronic interaction parameters is
transferable between a wide range of organic conjugated polymer
systems.

Another general consequence of localized electronic states in
molecular semiconductors is their effect on the molecule itself.
Promoting an electron from a $\pi$ bonding orbital to a $\pi^*$
anti-bonding orbital decreases the bond order over several
carbon-carbon bonds.  This leads to a significant rearrangement of the
bond-lengths to accommodate the changes in the electronic structure.
By in large, for polymers containing phenyl rings, it is the C=C bond
stretching modes and much lower frequency phenylene torsional modes
that play significant roles in the lattice reorganization following
optical excitation.  This is evidenced in the strong vibronic features
observed in the absorption and emission spectra of these materials.

Finally, one can fabricate devices using blends of semiconducting
polymers which phase segregate.  For example, the phenylene backbone
in F8BT is very planar molecule facilitating very delocalized
$\pi$-states.  On the other hand, TFB and PFB are very globular
polymers due to the tri-amide group in the chain.  Consequently, phase
segregation occurs due to more favorable $\pi$-stacking interactions
between F8BT chains than between F8BT and TFB or PFB.  Moreover, the
electronic states in TFB and PFB are punctuated by the tri-amides.
This difference in electronic states results in a band off-set between
the two semiconducting phases.  When we place the materials in contact
with each other, a {\em p-n} heterojunction forms.

\begin{figure}[t]
 \includegraphics[width=\columnwidth]{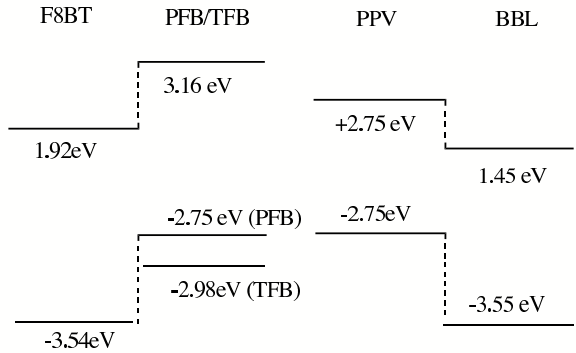}
 \caption{Relative placement of the HOMO and LUMO levels for various 
 conjugated polymers.}\label{bands}
 \end{figure}

One can think of the HOMO and LUMO energy levels of a given
polymer as corresponding to the top and bottom of the valance and
conduction bands respectively.  For the polymers under consideration
herein, the relative band edges are shown in Fig.~\ref{bands}.  In
Type II heterojunction materials, the energy bands of the two
materials are off set by $\Delta E$.  If the exciton binding energy
$\varepsilon_B > \Delta E$, excitonic states will the lowest lying
excited state species, resulting in a luminescent material with the
majority of the photons originating from the side with the lowest
optical gap.  Since the majority of the charge carriers are consumed
by photon production, very little photocurrent will be observed.

On the other hand, charge transfer states across the interface will be
energetically favored if $\varepsilon_B < \Delta E$.  Here, any
exciton formed will rapidly decay in to a charge-separated state with
the electron and hole localizing on either side of the junction.  This
will result in very little luminescence but high photocurrent.
Consequently, heterojunctions of PPV and BBL which have a large band
off-set relative to the exciton binding energy are excellent candidate
materials for organic polymer solar cells.

Heterojunctions composed of TFB:F8BT and PFB:F8BT lie much closer to
the exciton stabilization threshold as seen by comparing the relative
band off-sets in Fig~\ref{bands}.  Notice that the off-set for
TFB:F8BT is only slightly larger than 0.5eV, which is approximately
the exciton binding energy where as in PFB:F8BT the off-set is
$>0.5$eV.  Since such blends lie close to the stabilization threshold,
they are excellent candidates for studying the relation between the
energetics and the kinetics of exciton fission.

In this paper we will present an overview of our recent theoretical work 
aimed at understanding and modeling the state-to-state photophysical 
pathways in blended heterojunction materials.\cite{bittner:214719,photovolt2}
 
% In this paper we present an overview of our recent work in developing
% a dynamical model for electronic relaxation processes in molecular
% semiconductors.  We start with the a brief primer on the excited
% states of molecular semiconductors and develop concepts from solid
% state physics that are important in understanding molecular
% semiconductors.  We then provide details of a model we have developed
% over the past few years which captures much of the salient physics
% for the photophysics of molecular semiconductors with non-degenerate
% ground states, such as PPV, F8BT, and related polymers. 
% 
%  We will
% focus upon two interesting effects that affect the performance and
% operation of light-emitting diodes fabricated using organic polymers.
% First, the relative ratios of the singlet/triplet exciton capture
% cross-sections and secondly, the regeneration of secondary excitons
% in type II organic semiconductor heterojunctions.

\section{Two-band configuration interaction model}

Our basic description is derived starting from a model for the
on-chain electronic excitations of a single conjugated polymer
chain.\cite{karabunarliev:4291,karabunarliev:10219,karabunarliev:057402}
This model accounts for the coupling of excitations within the
$\pi$-orbitals of a conjugated polymer to the lattice phonons using
localized valence and conduction band Wannier functions
($|\overline{h}\>$ and $|p\>$) to describe the $\pi$ orbitals and two
optical phonon branches to describe the bond stretches and torsions of
the the polymer skeleton.

\begin{eqnarray}
H &=& \sum_{{\bf m n}} (F^\circ_{\bf mn} +V_{\bf mn} )A_{\bf
m}^\dagger A_{\bf n} \nonumber \\ &+& \sum_{{\bf nm} i\mu
}\left(\frac{\partial F^\circ_{\bf nm}}{\partial q_{i\mu}} \right)
A_{\bf n}^\dagger A_{\bf m} q_{i\mu} \nonumber \\ &+&
\sum_{i\mu}\omega_\mu^2(q_{i\mu}^2 + \lambda_\mu q_{i\mu} q_{i+1,\mu})
+ p_{i\mu}^2
\end{eqnarray}
where  $A^\dagger_{\bf n}$ and $A_{\bf n}$ are Fermion operators that act 
upon the ground electronic state $|0\rangle$ to
create and destroy electron/hole configurations
$|n\>  = | \overline{h}p \>$ with positive hole in the valence band Wannier function localized at 
$h$ and an electron in the conduction band Wannier function $p$.  Finally,  $q_{i\mu}$ and $p_{i\mu}$
correspond to lattice distortions and momentum components in the $i$-th site and $\mu$-th optical phonon branch.

Wannier functions are essentially spatially localized basis functions
that can be derived from the band-structure of an extended system.
Quantities such as the exchange interaction and Coulomb interaction
can be easily computed within the atomic orbital basis; however, there
are many known difficulties in computing these within the crystal
momentum representation.  Because of this, is is desirable to develop
a set of orthonormal spatially localized functions that can be
characterized by a band index and a lattice site vector, $R_\mu$.
These are the Wannier functions, which we shall denote by
$a_n(r-R_\mu)$ and define in terms of the Bloch functions
\begin{eqnarray}
a_n(r-R_\mu) = \frac{\Omega^{1/2}}{(2 \pi)^{d/2}}
\int e^{-ikR_\mu} \psi_{nk}(r) dk. 
\end{eqnarray}
The integral is over the Brillouin zone with volume $V= (2\pi)^d/\Omega$ and $\Omega$ is 
the volume of the unit cell (with $d$ dimensions). 
 A given Wannier function is defined for each band and for each unit cell.  
If the unit cell happens to contain multiple atoms, the Wannier function may be 
delocalized over multiple atoms.   The functions are orthogonal and complete.

The Wannier functions are not energy eigenfunctions of the
Hamiltonian.  They are, however, linear combinations of the Bloch
functions with different wave vectors and therefore different
energies.  For a perfect crystal, the matrix elements of $H$ in terms
of the Wannier functions are given by
\begin{widetext}
\begin{eqnarray} 
\int a^*_l(r-R_\nu) H_o a_n(r-R_\mu) dr &=& \frac{\Omega}{(2\pi)^d}
\int e^{i(qR_\nu-kR_\mu)}\psi_{lk}(r)H_o\psi_{nk}(r) dr dq dk\nonumber  \\
&=& {\cal E}_n(R_\nu-R_\mu)\delta_{nl}
\end{eqnarray}
where 
$$ {\cal E}_n(R_\nu-R_\mu) = \frac{\Omega}{(2\pi)^d}\int
 e^{ik(R_\nu-R_\mu)}E_n(k) dk.
 $$
 Consequently, the Hamiltonian matrix elements in the Wannier representation are 
related to the Fourier components of the band structure, $E_n(k)$.   Therefore, given
a band structure, we can derive the Wannier functions and the single particle 
matrix elements,  $F_{\bf mn}^\circ$.
\end{widetext}

\begin{figure}[t]
\begin{center}
\includegraphics{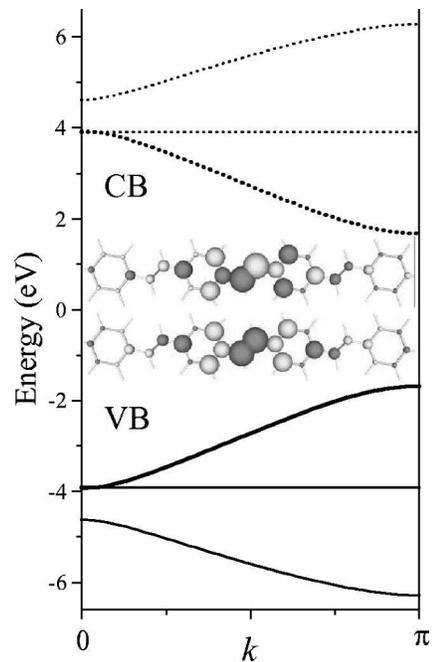}
\end{center}
\caption{Computed band-structure and vinylene-centered Wannier functions for PPV.}\label{wannier}
\end{figure}

The single-particle terms, $F_{\bf mn}^\circ$, are derived at the
ground-state equilibrium configuration, $q_\mu = 0$, from the Fourier
components $f_r$ and $\overline{f}_r$ of the band energies in
pseudomomentum space.
\begin{eqnarray}
F_{\bf mn} &=& \delta_{\overline{m}\overline{n}}\<m| f  |n\> - \delta_{mn}\<\overline{m}|\overline{f}  |\overline{n}\>\nonumber \\
&=&\delta_{\overline{m}\overline{n}}f_{m-n} - \delta_{mn}\overline{f}_{\overline{m}-\overline{n}}
\end{eqnarray}
Here, $f_{mn}$ and $\overline{f}_{mn}$ are the localized energy levels
and transfer integrals for conduction-band electrons and valence-band
holes.  At the ground-state equilibrium geometry, $q_\mu = 0$, these
terms can be computed as Fourier components of the one-particle
energies in the Brillouin zone.  For example, for the conduction band:
\begin{eqnarray}
f  _{mn} = f  _{m-n} = \frac{1}{B_z}\int_{B_z} \varepsilon_k   e^{ik(m-n)}dk
\end{eqnarray}
where $k$ is the pseudomomentum for a 1-dimensional lattice with unit period. 
For the case of cosine-shaped bands, 
$$
\epsilon(k) = f_{o} + 2 f_1\cos(k)
$$ the site-energies are given by the center $f_o$ and the transfer
integral between adjacent Wannier functions is given by $f_1$.  The
band-structure and corresponding Wannier functions for the valence and
conduction bands for PPV are shown in
Fig.~\ref{wannier}.\cite{karabunarliev:3988,karabunarliev:4291,karabunarliev:10219}
For the intrachain terms, we use the hopping terms and site energies
derived for isolated polymer chains of a given species, $t_{i,||}$,
where our notation denotes the parallel hopping term for the $i$th
chain ($i = 1,2$).  For PPV and similar conjugated polymer species,
these are approximately $0.5$eV for both valence and conduction $\pi$
bands.

The two-particle interactions are spin dependent with 
\begin{eqnarray}
V_{mn}^T = -\<m\overline{n}||n\overline{m}\> \\
V_{mn}^S = V_{mn}^T  + 2 \<m\overline{n}||\overline{m}n\>.
\end{eqnarray}
 for triplet and singlet combinations respectively with
$$ \<m\overline{n}||i\overline{j}\> = \int d1\int d2
\phi^*_m(1)\phi_{\overline{n}}^*(2)
v(12)\phi_i(1)\phi_{\overline{j}}(2)$$ With the exception of geminate
WFs, orbital overlap is small such that the two-body interactions are
limited to Coulomb, $J(r)$ and exchange integrals, $K(r)$ reflecting
e-h attraction and spin-exchange coupling non-geminate configurations.
and transition dipole-dipole integrals $D(r)$ coupling only geminate
singlet electron-hole pairs.  Table~\ref{params} gives a listing of
the electron-hole integrals and their parameters we have determined
for PPV and similar poly-phenylene based conjugated chains.  We have
found that these are quite transferable amongst this class of
conjugated polymers and allow us to focus upon modeling similar
poly-phenylene chains through variation of the Wannier function
band-centers (i.e. site-energies) and band-widths (i.e. intrachain
hopping integrals).

 \begin{table}[t]
 \begin{center}
  \caption{Electron/hole integrals for poly-phenylene-type polymer chains.}
 \begin{tabular}{c|c|c}
 \hline\hline
Term  &   Functional form  & Parameters \\
\hline
Direct Coloumb  & $J(r) = J_o/(1+r/r_o)$ & $J_o = 3.092 eV$ \\
                              &      				& $r_o =0.6840 a$ \\
\hline                              
Exchange           & $K(r) = K_oe^{-r/r_o}$ & $K_o = 1.0573 eV$ \\ & &  $r_o = 0.4743 a$ \\
\hline
Dipole-Dipole     &  $D(r) = D_o(r/r_o)^{-3}$ & $D_o = -0.03209 eV$ \\ & &  $r_o = 1.0a$\\
\hline
\hline
 \end{tabular}\label{params}
 \end{center}
Note: $a$ = unit lattice spacing. 
 \end{table}

Since we will be dealing with inter-chain couplings, we make the
following set of assumptions.  First, the single-particle coupling
between chains is expected to be small compared to the intramolecular
coupling.  For this, we assume that the perpendicular hopping integral
$t_\perp = 0.01 eV$.  This is consistent with LDA calculations
performed by Vogl and Campbell and with the $t_\perp \approx 0.15 f_1$
estimate used in an earlier study of interchain excitons by Yu {\em et
al}.\cite{vogl:12797,yu:8847} Furthermore, we assume that the $J(r)$,
$K(r)$, and $D(r)$ two-particle interactions depend only upon the
linear distance between two sites, as in the intrachain case.  Since
these are expected to be weak given that the interchain separation,
$d$, is taken to be some what greater than the inter-monomer
separation.

An important component in our model is the coupling between the
electronic and nuclear degrees of freedom. These we introduce via a
linear coupling term:
\begin{eqnarray}
\left( \frac{\partial f_{mn}}{\partial q_\mu}\right)_\circ = \frac{S}{2}(2\hbar \omega^3)^{1/2}
(\delta_{m\mu} + \delta_{n\mu})\label{huang}
\end{eqnarray}
where $S$ is the Huang-Rhys factor which can be obtained from vibronic features in
the experimental photoemission spectra. For the case of conjugated polymers such as PPV 
and similar poly-phenylene species, the emission spectra largely consists of 
a series of well-resolved vibronic features corresponding to the C=C stretching modes 
in the phenylene rings with typical Huang Rhys factors of 0.4.  In addition, low frequency
vibrational modes (due to torsions) contribute a broad featureless background.  
Consequently.  we include two intramolecular optical phonon branches which correspond roughly to 
the high-frequency C=C bond stretching modes within a given repeat unit and a second low-frequency
mode, which in the case of PPV are taken to represent the phenylene torsional modes.  
The electron-phonon couplings are assumed to be transferable between the various chemical species. 
Since the modes are assumed to be intramolecular, we do not include interchain couplings
in the phonon Hessian matrix.

Upon transforming $H$ into the diabatic representation by
diagonalizing the electronic terms at $q_{i\mu} = 0$, we obtain a
series of vertical excited states $|a_\circ\>$ with energies,
$\varepsilon_a^\circ$ and normal modes, $Q_\xi$ with frequencies,
$\omega_\xi$.  (We will assume that the sum over $\xi$ spans all
phonon branches).
\begin{eqnarray}
H &=& \sum_a \varepsilon_a^\circ |a_\circ\> \<a_\circ|+
\sum_{ab\xi}g^\circ_{ab\xi}q_\xi(|a_\circ\>\<b_\circ|
+|b_\circ\>\<a_\circ| ) \nonumber \\ &+& \frac{1}{2}\sum_\xi
(\omega_\xi^2 Q_\xi^2 + P_\xi^2 ).\label{diab}
\end{eqnarray}
The adiabatic or relaxed states can be determined then by iteratively
minimizing $\varepsilon_a(Q_\xi) = \<a|H|a\>$ according to the
self-consistent equations
\begin{eqnarray}
\frac{d\varepsilon_a(Q_\xi) }{dQ_\xi}  = g_{aa\xi} + \omega_\xi^2Q_\xi  = 0.
\end{eqnarray}

\begin{figure}[t]
\includegraphics[width=\columnwidth]{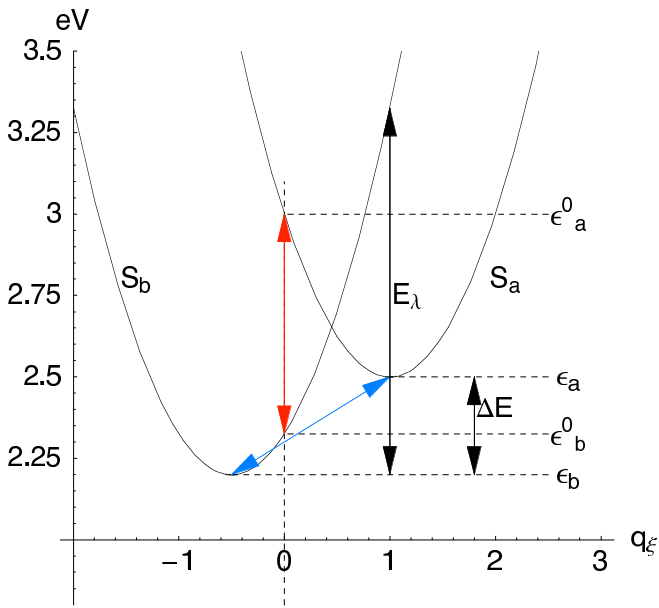}
\caption{Schematic representation of excited state Diabatic potentials
obtained within our approach. The ground state configuration is taken
as $q_\xi = 0$ with vertical excitation energies at$\epsilon_a^\circ$
and $\epsilon_b^\circ$ and adiabatic (minimum) energies at
$\epsilon_a$ and $\epsilon_b$. }\label{marcus}
\end{figure}

Thus, each diabatic potential surface for the nuclear lattice motion is given by
\begin{eqnarray}
\varepsilon_a(Q_\xi) = \varepsilon_a + \frac{1}{2}\sum_\xi
\omega_\xi^2 (Q_\xi - Q_{\xi}^{(a)})^2.
\end{eqnarray}
These are shown schematically as $S_a$ and $S_b$ in Fig.~\ref{marcus}
with $Q_\xi$ being a collective normal mode coordinate.  On can also
view this figure as a slice through an $N-$dimensional coordinate
space along normal coordinate $Q_\xi$ In this figure,
$\epsilon_a^\circ$ and $\epsilon_b^\circ$ are the vertical energies
taken at the ground-state equilibrium geometry $Q_\xi = 0$.  The
adiabatic energies, taken at the equilibrium geometry of each excited
state are denoted as $\epsilon_a$ and $\epsilon_b$.  While our model
accounts for the distortions in the lattice due to electron/phonon
coupling, we do not account for any adiabatic change in the phonon
force constants within the excited states.  Lastly, the electronic
coupling between diabatic curves is given by $g_{ab}$ which we compute
at the ground-state geometry ($g_{ab}^\circ$).  We assume that both
the diagonal $g_{aa}^\circ$ and off-diagonal $g_{ab}^\circ$ terms can
be derived from the spectroscopic Huang-Rhys parameters as in
Eq.~\ref{huang}.

The advantages of our approach is that it allows us to easily consider
the singly excited states of relatively large conjugated polymer
systems.  Our model is built from both {\em ab initio} and
experimental considerations and can in fact reproduce most of the
salient features of the vibronic absorption and emission spectra for
these systems.  The model is limited in that we cannot include
specific chemical configurational information about the polymers other
than their conjugation length and gross topology.  For isolated single
chains, the model is rigorous.  For multiple chains, our interchain
parameterization does not stand on such firm ground since technically
the Wannier functions are derived from a quasi-one-dimensional band
structure.  Nonetheless, our model and results provide a starting
point for predicting and interpreting the complex photophysical
processes within these systems.  We next move on to describing the
state-to-state interconversion proceses that occur following both
photo and electro-excitation.

%\includegraphics[width=\columnwidth]{10-10-Lattice}

%The best verification of our model is direct comparison to the experimental photo-absorption and emission  spectra for various systems.  Here, we show the experimental and predicted 
%spectra for F8BT.  
%\begin{figure}[h]
%\includegraphics[width=4.0in]{PF-spectra}
%\includegraphics[width=4.0in]{f8bt-spectra-bw}
%\caption{Theoretical and Experimental F8BT spectra shows "double peaked" absorption spectra and 
%single band emission.}
%\end{figure}
%\begin{figure}[b]
%\includegraphics[width=\columnwidth]{f8bt-spectra-bw}
%\caption{ Absorption (dashed) and vibronic 
%emission (solid) spectra for the modulated F8BT chain. The vibronic fine-structure
%in each peak is due to the C=C bond-stretching phonon branch.} \label{f8bt-mod-spectrum}
%\end{figure}

\section{State to State relaxation Dynamics}
The electronic levels in our model are coupled to the lattice phonons
as well as the radiation field.  Consequently, relaxation from a given
electronic state can occur via state to state interconversion via
phonon excitation or absorption or fluorescent decay to the $S_o$
ground state.  For triplet excitations, only phonon transitions are
allowed.  For the singlets, fluorescence occurs primarily from the
lowest $S_n$ state independent of how the excitation was prepared.
This certainly holds true for conjugated polymers in which both
electroluminescence and photoluminescence originates from the same
$S_1 = S_{xt}$ state.  This implies that internal conversion dynamics
are fast relative to the fluorescence lifetime.

Coupling the electronic relaxation dynamics to the vibrational
dynamics is a formidable task.  An exact quantum mechanical
description of this is currently well beyond the state of the art of
current computational methods.  One can, however, compute the
state-to-state rate constants using Fermi's golden rule and arrive at
a reasonable picture.

If we assume that the vibrational bath described by $H_{ph}$ remains
at its ground-state geometry, then the state-to-state transition rates
are easily given by Fermi's golden rule:
\begin{eqnarray}
k_{ab}^\circ  = \pi \sum_\xi \frac{g^2_{ab}}{\hbar \omega_\xi} (1 + n(\omega_{ab}) ) 
(\Gamma(\omega_\xi - \omega_{ab}) - \Gamma(\omega_\xi + \omega_{ab})).\label{static}
\end{eqnarray}
where $n(\omega)$ is the Bose-Einstein population for the phonons,
$\Gamma$ is a Lorentzian broadening in which the width is inversely
proportional to the phonon lifetime used to smooth the otherwise
discrete phonon spectrum, $\omega_{ab} = (\epsilon_a^o -
\epsilon_b^o)/\hbar$.  In order for a transition to occur, there must
be a phonon of commensurate energy to accommodate the energy transfer.
The coupling term, $g_{ab\xi}$, is the diabatic coupling in the
diabatic Hamiltonian given in Eq.~\ref{diab}.

This static model is fine so long as either the nuclear relaxation has
little effect on the state to state rate constant or if the electronic
transitions occur on a time-scale which is short compared to the
nuclear motion.  However, if lattice reorganization does play a
significant role, then we need to consider the explicit nuclear
dynamics when computing the state-to-state rates.  If we assume that
vibrational relaxation within a given diabatic state is rapid compared
to the inter-state transition rate, we can consider the transitions as
occurring between displaced harmonic wells
\begin{eqnarray}
k_{ab} = \frac{2\pi}{\hbar} \left| V_{ab}\right|^2 {\cal F}
\end{eqnarray}
where $V_{ab}$ is the coupling between electronic states $a$ and $b$ and 
\begin{eqnarray}
{\cal F}(E_{ab}) &=& \sum_{\nu_a}\sum_{\nu_b} P_{th}(\varepsilon_a(\nu_a))
|\langle \nu_a | \nu_b\rangle|^2 \nonumber \\
&\times&\delta(\epsilon_a(\nu_a) - \epsilon(\nu_b) + \Delta E_{ab})
\end{eqnarray}
is the thermally averaged Franck-Condon weighted density of nuclear
vibrational states.  Here, $\nu_a$ and $\nu_b$ denote the vibronic
states, $P_{th}$ is the Boltzmann distribution over the initial
states, $\epsilon_a(\nu_a)$ and $\epsilon(\nu_b)$ are the
corresponding enregies, and $\Delta E_{ab}$ is the electronic energy
gap between $a$ and $b$.  In the classical limit, ${\cal F}$ becomes
\begin{eqnarray}
{\cal F}(E_{ab})   = \frac{1}{\sqrt{4 \pi E_\lambda k_B T}}
\exp\left(-\frac{(E_\lambda +\Delta E_{ab})^2}{4 E_\lambda k_B T} \right)
\end{eqnarray}
where $E_\lambda$ is the reorganization energy as sketched in Fig.~\ref{marcus}.

Each of these terms can be easily computed from the diabatic
Hamiltonian in Eq.~\ref{diab}.  The diabatic coupling matrix element
between the adiabatically relaxed excited states, $|V_{ab}|^2$,
requires some care since we are considering transitions between
eigenstates of different Hamitonians (corresponding to different
nuclear geometries).  Since the vertical $Q_\xi = 0$ states provide a
common basis, $|a^\circ\rangle$, we can write
\begin{eqnarray}
V_{ab} = \sum_{a^\circ b^\circ} \langle a | a^\circ\rangle
g_{ab}^\circ \langle b^\circ | b\rangle
\end{eqnarray}
where $g_{ab}^\circ$ is the diabatic matrix element computed at the
equilibrium geometry of the ground-state.

Once we have the rate constants computed, it is a simple matter to
integrate the Pauli master equation for the state populations.
\begin{eqnarray}
\dot{P}_a(t) = \sum_{b} \left(k_{ba}P_b-k_{ab}P_a\right)  - k^{rad}_a P_a
\end{eqnarray}
where $k^{rad}_a$ is the radiative decay rate of state $a$.
\begin{eqnarray}
k_{a}^{rad} = \frac{|{\bf
\mu}_{a0}|^2}{6\epsilon_o\hbar^2}(1+n(\omega_{a0}))\frac{\hbar
\omega_{a0}^3}{2\pi c^3}
\end{eqnarray}
where ${\bf \mu}_{a0}$ are the transition dipoles of the excited
singlets.  These we can compute directly from the Wannier functions or
empirically from the photoluminescence decay rates for a given system.
Photon mediated transitions between excited states are highly unlikely
due to the $\omega_{ab}^3$ density factor of the optical field.  In
essence, so long as the non-equilibrium vibrational dynamics is not a
decisive factor, we can use these equations to trace the relaxation of
an electronic photo- or charge-transfer excitation from its creation
to its decay including photon outflow measured as luminescence.

\section{Exciton Regeneration Dynamics}

Donor-acceptor heterojunctions composed of blends of TFB with F8BT and
PFB with F8BT phase segregate to form domains of more or less pure
donor and pure acceptor.  Even though the polymers appear to be
chemically quite similar, the presence of the tri-phenyl amine groups
in TFB and PFB cause the polymer chain to be folded up much like a
carpenter's rule.  F8BT, on the other hand, is very rod-like with a
radius of gyration being more or less equivalent to the length of a
give oligomer.  Molecular dynamics simulations of these materials by
our group indicate that segregation occurs because of this difference
in morphology and that the interface between the domains is
characterized by regions of locally ordered $\pi$-stacking when F8BT
rod-like chains come into contact with more globular PFB or TFB
chains.

As discussed earlier, TFB:F8BT and PFB:F8BT sit on either side of the
exciton destabalization threshold.  In TFB:F8BT, the band off-set is
less than the exciton binding energy and these materials exhibit
excellent LED performance.  On the other hand, devices fabricated from
PFB:F8BT where the exciton binding energy is less than the off-set,
are very poor LEDs but hold considerable promise for photovoltic
devices.  In both of these systems, the lowest energy state is assumed
to be an interchain exciplex as evidenced by a red-shifted emission
about 50-80ns after the initial photoexcitation.  \cite{morteani:1708}
In the case of TFB:F8BT, the shift is reported to be 140$\pm$20 meV
and in PFB:F8BT the shift is 360$\pm$30 meV relative to the exciton
emission, which originates from the F8BT phase.  Bearing this in mind,
we systematically varied the separation distance between the cofacial
chains from $r = 2a - 5a$ (where $a$ = unit lattice constant) and set
$t_\perp = 0.01$ in order to tune the Coulomb and exchange coupling
between the chains and calibrate our parameterization.

 For large interchain separations, the exciton remains localized on
 the F8BT chain in both cases.  As the chains come into contact,
 dipole-dipole and direct Coulomb couplings become significant and we
 begin to see the effect of exciton destabilization.  For TFB:F8BT, we
 select and interchain separation of $r = 2.8a$ giving a 104meV
 splitting between the vertical exciton and the vertical exciplex and
 87.4 meV for the adiabatic states.  For PFB:F8BT, we chose $r = 3a$
 giving a vertical exciton-exciplex gap of 310 meV and an adiabatic
 gap of 233 meV. In both TFB:F8BT and PFB:F8BT, the separation produce
 interchain exciplex states as the lowest excitations.  with energies
 reasonably close to the experimental shifts.

\begin{figure}
\includegraphics[width=\columnwidth]{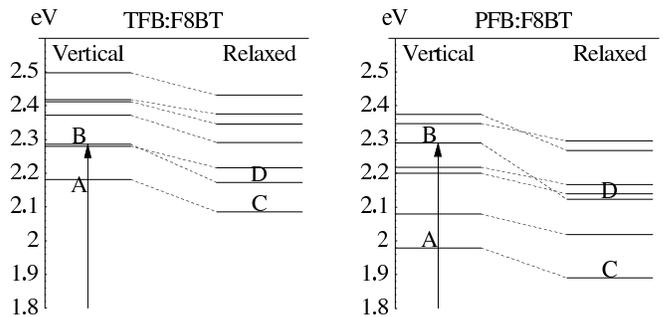}\label{tfb-pfb-f8bt-correlation}
\caption{Vertical and relaxed energies of the lowest lying states in
the TFB:F8BT and PFB:F8BT heterojunctions.  In each, A and C refer to
the interchain exciplex state and B and D refer to the predominantly
intrachain F8BT exciton state.}
\end{figure}

\begin{figure}[t]
\includegraphics[width=\columnwidth]{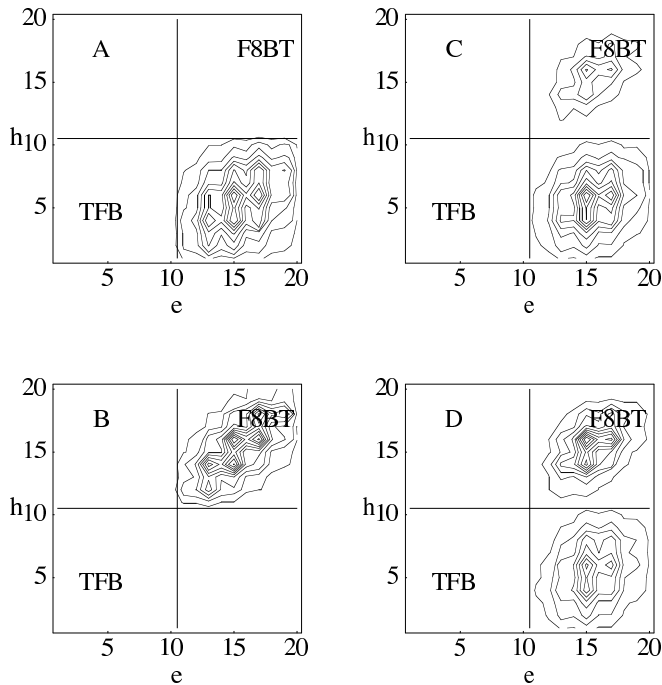}
\caption{
Excited state electron/hole densities for TFB:F8BT heterojunction.  
The electron/hole coordinate axes are such
 that  sites 1-10 correspond to TFB sites and 11-20 correspond to F8BT sites.
 Note the weak mixing between the 
interchain charge-separated states and the F8BT exciton in each of these plots.
}\label{tfb-f8bt-states}
\end{figure}

\begin{figure}[t]
\includegraphics[width=\columnwidth]{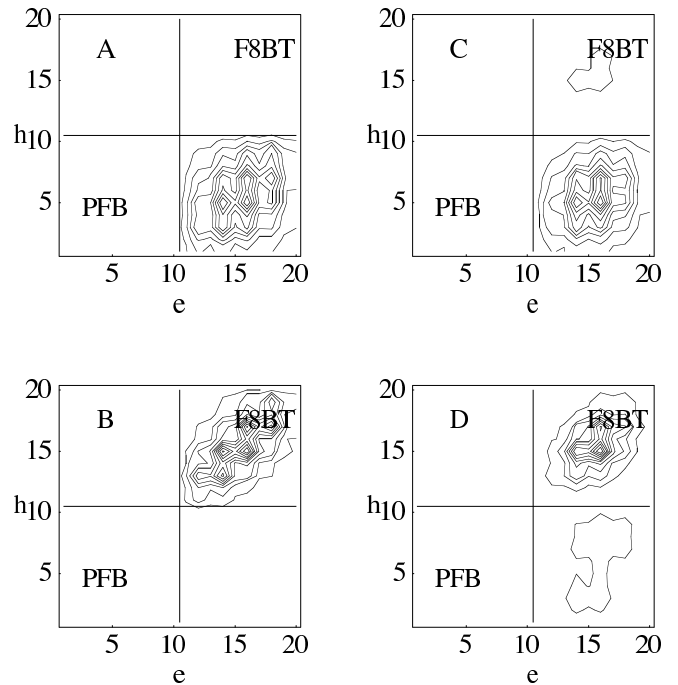}
\caption{
Excited state electron/hole densities for PFB:F8BT heterojunction.
The axes are as in previous figures  except that sites 1-10 correspond to PFB
sites and sites 11-20 to F8BT sites. }\label{pfb-f8bt-states}
\end{figure}

Fig.~\ref{tfb-pfb-f8bt-correlation} compares the vertical and
adiabatic energy levels in the TFB:F8BT and PFB:F8BT chains and
Figs.~\ref{tfb-f8bt-states} and \ref{pfb-f8bt-states} show the
vertical and relaxed exciton and charge-separated states for the two
systems.  Here, sites 1-10 correspond to the TFB or PFB chains and
11-20 correspond to the F8BT chain.  The energy levels labeled in
Fig.~\ref{tfb-pfb-f8bt-correlation} correspond to the states plotted
in Figs.~\ref{tfb-f8bt-states} and \ref{pfb-f8bt-states}.  We shall
refer to states A and B as the vertical exciplex and vertical exciton
and to states C and D as the adiabatic exciplex and adiabatic exciton
respectively.  Roughly, speaking a pure exciplex state will have the
charges completely separated between the chains and will contain no
geminate electron/hole configurations.  Likewise, strictly speaking, a
pure excitonic state will be localized to a single chain and have only
geminate electron/hole configurations.  Since site energies for the
the F8BT chain are modulated to reflect to internal charge-separation
in the F8BT co-polymer as discussed above, we take our ``exciton'' to
be the lowest energy state that is localized predominantly along the
diagonal in the F8BT ``quadrant''.

In the TFB:F8BT junction, the lowest excited state is the exciplex for
both the vertical and adiabatic lattice configurations with the hole
on the TFB and the electron on the F8BT.
(Fig.\ref{tfb-f8bt-states}A,C) In the vertical case, there appears to
be very little coupling between intrachain and interchain
configurations.  However, in the adiabatic cases there is considerable
mixing between intra- and inter-chain configurations.  First, this
gives the adiabatic exciplex an increased transition dipole moment to
the ground state.  Secondly, the fact that the adiabatic exciton and
exciplex states are only 87 meV apart means that at 300K, about 4\% of
the total excited state population will be in the adiabatic exciton.

For the PFB:F8BT heterojunction, the band off-set is greater than the
exciton binding energy and sits squarely on the other side of the
stabilization threshold.  Here the lowest energy excited state
(Figs.~\ref{pfb-f8bt-states}A and B) is the interchain
charge-separated state with the electron residing on the F8BT (sites
11-20 in the density plots in Fig~\ref{pfb-f8bt-states}) and the hole
on the PFB (sites 1-10).  The lowest energy exciton is almost
identical to the exciton in the TFB:F8BT case.  Remarkably, the
relaxed exciton (Fig.~\ref{pfb-f8bt-states}D) shows slightly more
interchain charge-transfer character than the vertical exciton
(Fig.~\ref{pfb-f8bt-states}C).  While the system readily absorbs at
2.3eV creating a localized exciton on the F8BT, luminescence is
entirely quenched since all population within the excited states is
readily transfer to the lower-lying interchain charge-separated states
with vanishing transition moments to the ground state.

\begin{figure}[t]
\includegraphics[width=\columnwidth]{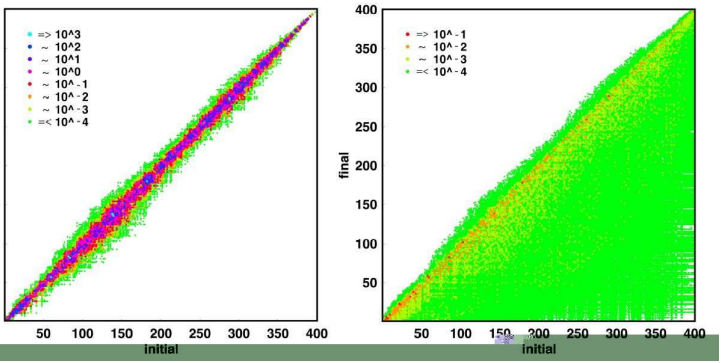}
	\caption{(Color online) TFB-F8BT internal conversion rates
	distribution at 290~K.  Rates are in ps$^{-1}$.  Note the
	sparsity and relatively faster Marcus rates compared to the
	diabatic rates.}
	\label{tfb-f8bt_allrates}
\end{figure}

In calculating the state-to-state interconversion rates for TFB:F8BT,
we note two major differences between the static and the adiabatic
Marcus-Hush approaches (See Fig.~\ref{tfb-f8bt_allrates}).  First is
the sparsity of the latter with transitions being limited to states
with smaller energy differences.  This leads to a relaxation dynamics
that is more intertwined with the DOS.  Second is the relatively
faster rates calculated in the latter leading to interconversion
lifetimes in the femtosecond (fs) to a couple of picosecond (ps)
regime as opposed to hundreds of ps in the former.  The same general
difference is observed for PFB:F8BT (Not shown).  These marked
differences in the distribution of rates and their range of magnitudes
are brought about by the introduction of the reorganization energy as
a parameter in the rates calculation to complement the energy
differences between the states.  It provides a way to incorporate
lattice distortions in the semiclassical limit into the relaxation
dynamics.  While this is not fully dynamical in its account of the
lattice distortions, it improves upon the static approximation
previously employed.

\begin{figure}
	\includegraphics[width=0.9 \columnwidth]{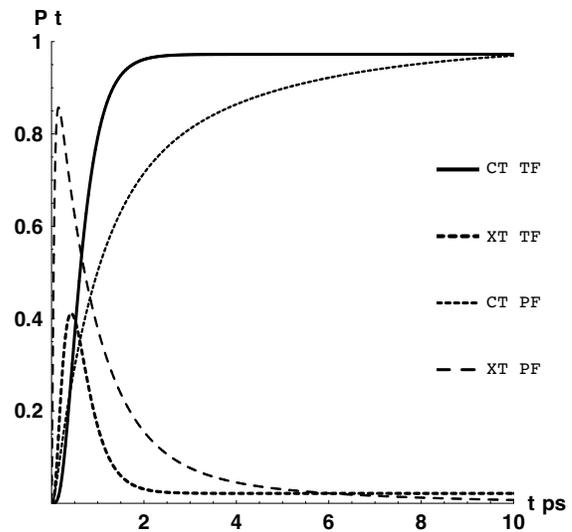}
	\caption{Time-evolved populations of the lowest CT (solid
	lines) and XT (dashed lines) relaxed states of TFB:F8BT(TF)
	and PFB:F8BT(PF) in the Marcus-Hush approach at 290~K.}
	\label{popevo_marc}
\end{figure}

The photoexcitation of heterojunction systems is simulated by
populating a higher-lying excitonic state.  Fig. \ref{popevo_marc}
shows the time-evolved populations of the lowest charge-transfer (CT)
and excitonic (XT) vertical and relaxed states, respectively, in
photoexcited TFB:F8BT and PFB:F8BT.  We see, in both approaches, that
the relaxation to the lowest CT state is faster in TFB:F8BT than in
PFB:F8BT.  Furthermore, the relaxation from the XT state to the CT
state occurs faster in the former.  This is despite the XT state being
formed faster in the latter for both cases.  This is manifested more
in the Marcus-Hush approach shown in Fig.~\ref{popevo_marc} where
despite reaching a maximum population of 0.86 in 250~fs as opposed to
just 0.40 in 500~fs, the XT$\rightarrow$CT interconversion is
practically done in 2~ps in TFB:F8BT compared to 10~ps in PFB:F8BT.
In addition, we note that the XT state reaches a steady state
population in TFB:F8BT whereas it goes to zero in PFB:F8BT.  This
small but non-zero population of the XT state is consistent with the
distributed thermal population of states of 0.022 at 290~K owing to
the fact that this XT state is 95~meV higher in energy relative to the
lowest CT state.\cite{bittner:214719}

        % Figure: Rates Distribution

\begin{figure}[h]
\includegraphics[width=0.45 \columnwidth]{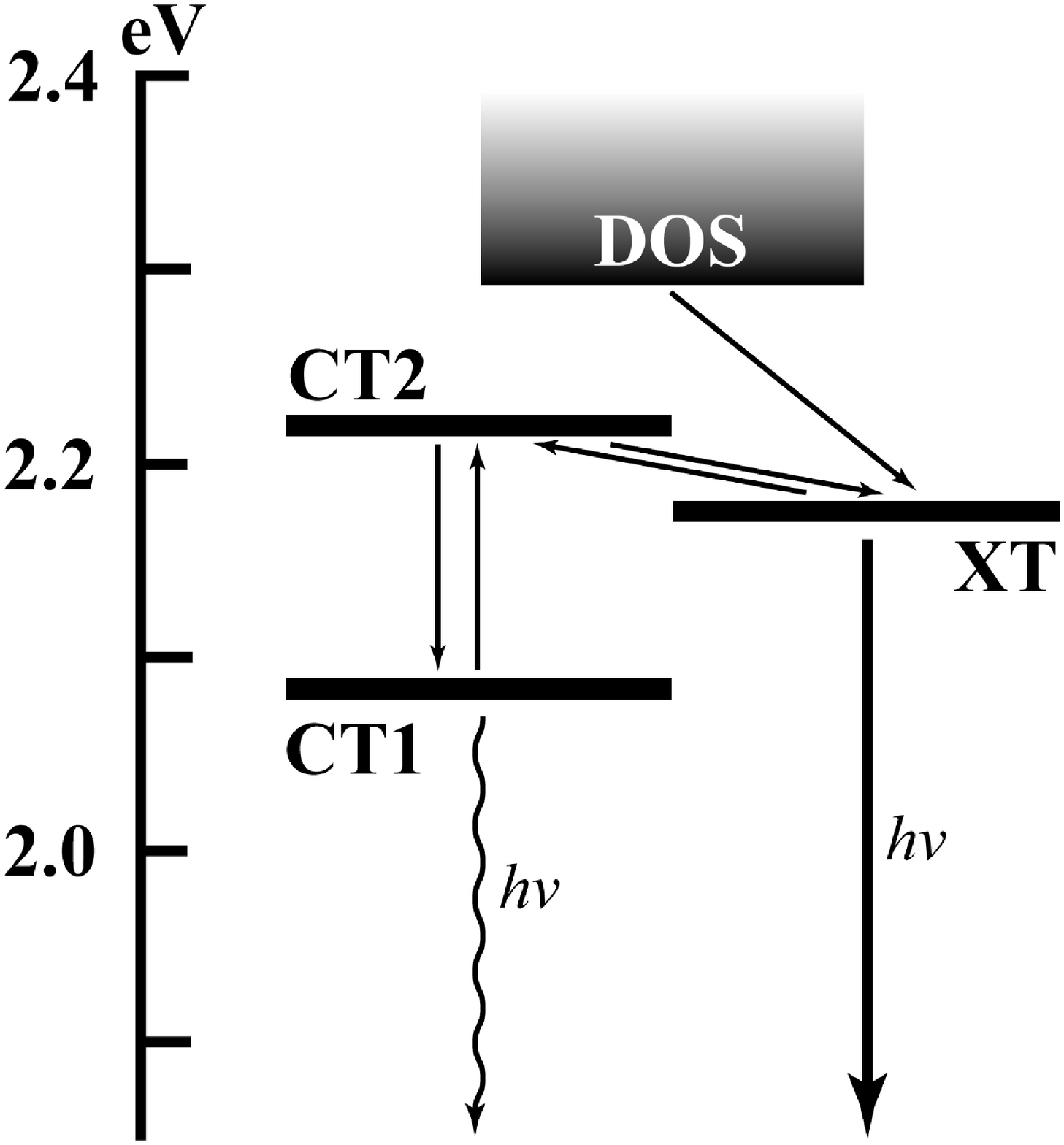}
\includegraphics[width=0.45 \columnwidth]{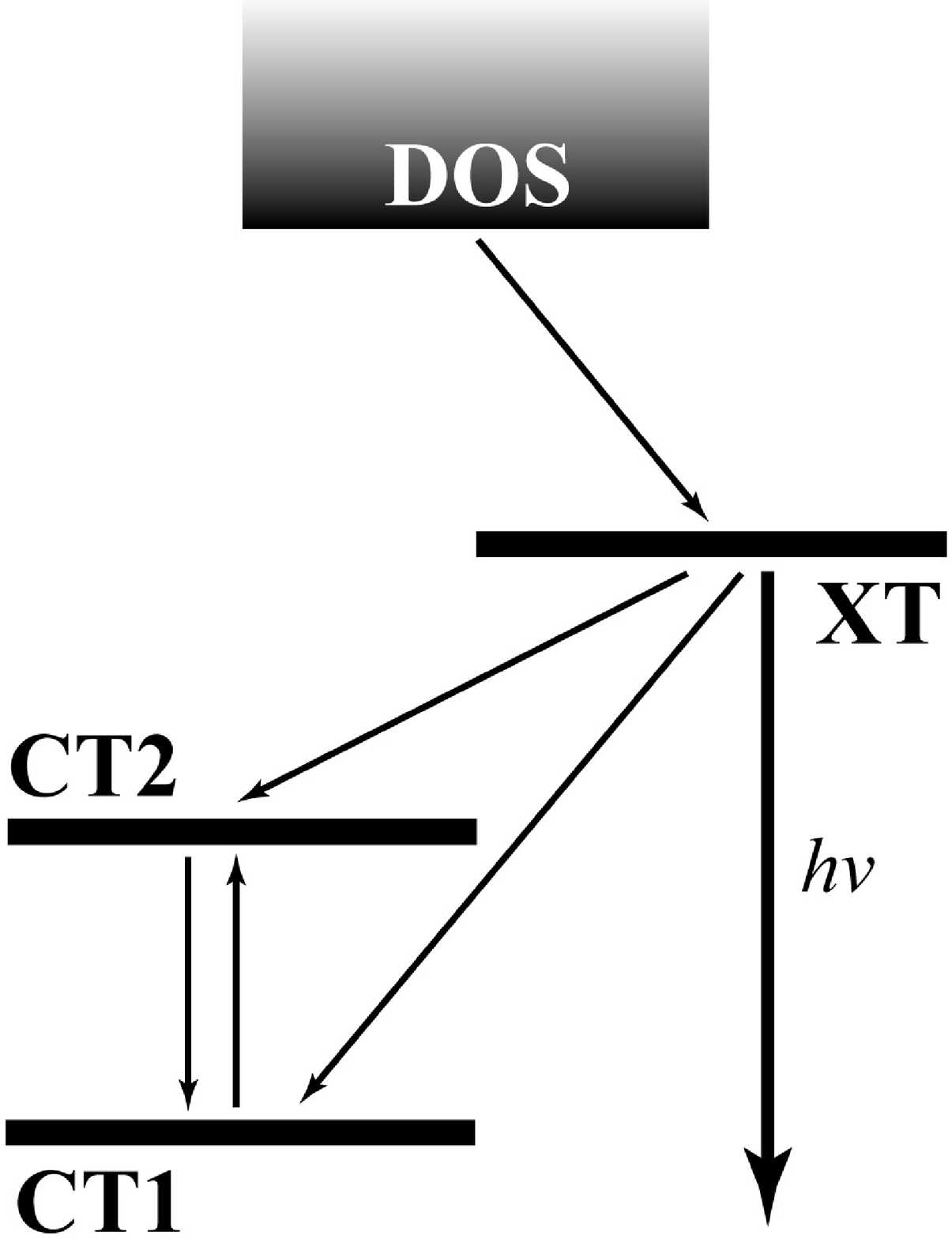}
	\caption{Relevant Marcus-Hush interconversion rates for the 3
	lowest states of (left)~TFB:F8BT and (right)~PFB:F8BT.  In
	both cases, relaxation proceeds from the density of states
	(DOS) to the lowest excitonic state(XT) (off-set to the right
	relative to the CT states for clarity of relaxation route)
	before relaxing to the lowest charge-transfer state(CT1).  CT1
	proceeds to equilibrium with the next higher CT state(CT2).
	In PFB:F8BT, CT2 has a lower energy than XT where as in
	TFB:F8BT, it has a higher energy.  Also shown are the
	radiative rates emanating from the XT states which are
	strongly coupled to ground state, S$_0$, of both systems and
	the TFB:F8BT CT1 state which is just weakly coupled to S$_0$.}
	\label{ic_low3}
\end{figure}

Interestingly, while the overall XT$\rightarrow$CT interconversion
occurs in just a couple of ps in both heterojunction systems, a closer
look into the rates reveal that this relaxation does not occur
directly.  Rather, it involves the next lowest CT state.
Fig.~\ref{ic_low3} show the relevant interconversions between the
three lowest states of both systems: the lowest CT state(CT1), the
next lowest CT state(CT2), and the lowest XT state(XT).
%Fig.~\ref{ehdens_low3} shows the electron/hole densities of these three lowest states for TFB:F8BT and PFB:F8BT.  
It is worth noting that the considerable mixing between the
intra-chain and inter-chain configurations of the former compared to
those of the latter.  In TFB:F8BT, the direct XT$\rightarrow$CT1
transition ($\sim 10^{-3}$~ps$^{-1}$) is at least 3 orders of
magnitude slower than the corresponding
XT$\rightarrow$CT2$\rightarrow$CT1 transition route $(>~1$~ps$^{-1}$),
the indirect route being consistent with the evolution data
(Fig.~\ref{popevo_marc}).  Thus, the XT$\rightarrow$CT conversion
occurs via the CT2 state and not directly.  The reverse transitions
for both routes are slower but have the same order of magnitude
difference.  This CT1$\rightarrow$CT2$\rightarrow$XT transition ($\sim
10^{-1}$~ps$^{-1}$) effectively presents a regeneration pathway for
the XT.  This leads to an XT state population that is always at
equilibrium with the CT1 state.

In PFB:F8BT the XT$\rightarrow$CT1 and XT$\rightarrow$CT2 conversion
occur at relatively the same rate ($\sim 10^{-1}$~ps$^{-1}$) while
their reverse transitions are at least 2 orders of magnitude slower.
Consequently, XT is not regenerated.  The role played by CT2 as a
bridge state is apparently relative to whether it has a slightly
higher or lower energy than the XT as has been accounted by Morteani
$et~al.$\cite{morteani:247402, morteani:1708}.  Spontaneous transition
rates are typically faster when going from a higher to a lower energy
state than the reverse according to detailed balance.  Here, CT1 is
the exciplex state which exhibit sizable mixing with the bulk CT state
(CT2).  When CT2 has a higher energy than XT, such as in TFB:F8BT, a
fraction of the population in CT2 converts to XT.  If it has a lower
energy relative to the XT state such as in PFB:F8BT, this regeneration
of the XT, practically, does not occur.

% Figure: Marcus-Hush population evolution

% Figure: Summary of Interconversions (3 lowest states)

\begin{figure}[floatfix]
	\includegraphics[width=\columnwidth]{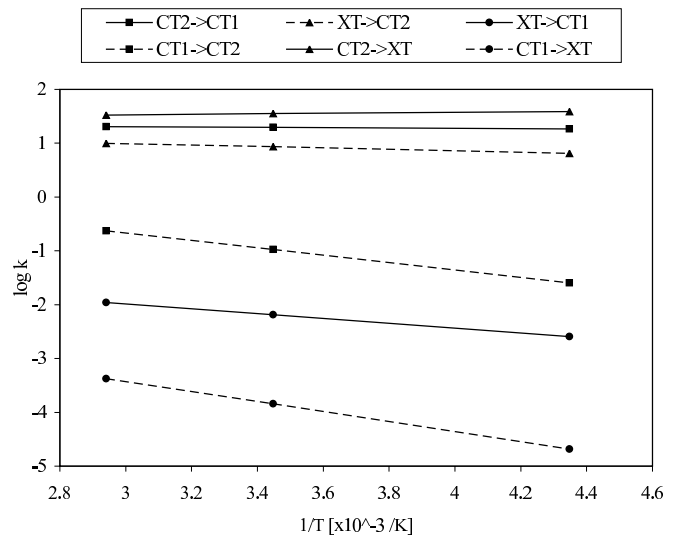}
	\caption{Interconversion rates between the 3 lowest states of
	TFB:F8BT as a function of temperature (230, 290, and 340~K).
	Plot is given as log~k $vs$ 1/T.  Transitions to lower energy
	states are given as solid lines while their reverse are given
	as dashed lines.  The CT2$\leftrightarrow$CT1,
	CT2$\leftrightarrow$XT and XT$\leftrightarrow$CT1 are plotted
	as squares, triangles and circles, respectively.  All
	transition rates increase directly with temperature except the
	CT2$\rightarrow$XT conversion which decreases as temperature
	increases.}
	\label{tfb-f8bt_tdepr8s}
\end{figure}

To see the effect of temperature, the interconversion rates were
calculated at 230, 290, and 340~K.  Fig.~\ref{tfb-f8bt_tdepr8s}
shows how the interconversions among the three lowest states (CT1,
  CT2, and XT) of TFB:F8BT, as illustrated in Fig.~\ref{ic_low3}, vary
  with temperature.  This dependence is given in an Arrhenius plot of
  log~k $vs$ 1/T and gives a linear plot for each transition having a
  slope associated with the activation energy, $E_{act}$, for that
  particular transition.  This activation energy has the expression,
\begin{equation}
	E_{act} = \frac{(\Delta E - 	E_\lambda)^2}{4E_\lambda}.
\end{equation}
  Transitions to lower energy states are given as solid lines while
those going to higher energy states are given as dashed lines.
Curiously, although XT$\rightarrow$CT1 is exothermic compared to
XT$\rightarrow$CT2 which is endothermic, the latter is a more
favorable transition.  This has to do with the fact that
XT$\rightarrow$CT1 has an activation energy almost three times greater
than that of XT$\rightarrow$CT2.  As alluded to above, this is a
consequence of the former being in the inverted region while the
latter being in the normal region.  In the inverted region, the larger
$\Delta E$ is, the larger $E_{act}$ as opposed to the more familiar
normal region where $E_{act}$ decreases as $\Delta E$ increases.
Having stated this, however, we note that in the former, due to
maximal overlap between the vibrational modes of the two states,
transitions may be possible via tunneling processes.  Overall, in
TFB:F8BT, we see an increase in the fraction of the total excited
state population in XT as temperature increases.  At 230~K only 0.81\%
is in XT while at 290 and 340~K, 2.16\% and 3.67\% is in XT,
respectively.

% Figure:  TFB:F8BT interconversion temperature dependence

Finally, we note that all transition rates increase with temperature
 except for the CT2$\rightarrow$XT in TFB:F8BT and XT$\rightarrow$CT2
 in PFB:F8BT which decrease with temperature.  Such a trend, while not
 uncommon in chemical reactions, are though to be indicative of a more
 complicated transition mechanism as noted by Porter\cite{porter:241}.
 We surmise this to be due to the coupling between the low frequency
 vibrational modes of the initial state with the high frequency
 vibrational modes of the final state as in the case of an early
 transition state in reactive scattering.  What we are definite of is
 that both transitions are exothermic and in the normal region which
 makes one speculate if there is any connection.

\section{Discussion}
In this paper, we gave an overview of our recent work in developing a
theoretical understanding interfacial excitonic dynamics in a complex
material system.  The results herein corroborate well with the
experimental results on these systems.  In particular, following
either charge injection or photo-excitation, the system rapidly
relaxes to form the interchain charge-separated species.  In the
experimental data, this occurs within the first 10 or so ps for the
bulk material.  Our calculations of a single pair of cofacial chains
puts the exciplex formation at about 1 ps.  Likewise, the experimental
time-resolved emission indicates the regeneration occurs on a much
longer time-scale with most of the time-integrated emission coming
from regenerated excitons.  This too, is shown in our calculations as
evidenced by the slow thermal repopulation of the XT state in the
TFB:F8BT system.  Since this state has a significant transition dipole
to the ground state, population transfered to this state can either
decay back to the CT state via thermal fluctuations or decay to the
ground state via the emission of a photon.  Since this secondary
emission is dependent upon the thermal population of the XT state at
any given time, the efficiency of this process shows a strong
dependency upon the temperature of the system.

The exciton model we present herein certainly lacks the molecular
level of details so desired by materials chemists.  However, it offers
a tractable way of building from molecular considerations the salient
physical interactions that give rise to the dynamics in the excited
states of these extended systems.  In building this model we make a
number of key assumptions.  First, and perhaps foremost, that the
excited states are well described via bands of $\pi$ orbitals and that
from these bands we can construct localized Wannier functions. Hand in
hand with this assumption is that within the general class of
oligo-phenylene derived polymers, configuration interaction matrix
elements, hopping integrals, electron/phonon couplings, and phonon
spectra are transferable from one system to another.  This is a fairly
dangerous approximation since it discounts important contributions
from heteroatoms, side-chains, and chain morphology.  However, given
that a single oligomeric chain of F8BT with 10 repeat units has well
over 300 atoms, such potentially dire approximations are necessary in
order to extract the important features of these very extended
systems.

Secondly, we make the assumption that the explicit vibrational
dynamics can be integrated out of the equations of motion for the
electronic states.  This is probably not too extreme of an assumption
so long as we can assume that the phonons remain thermalized over the
course of the electronic relaxation.  However, looking back at the
level correlation diagrams, crossings between diabatic states are
present in this system and hence conical intersections between
electronic states may play an important role.  Finally, we discount
the effects of electronic coherence.  This, too, may have a profound
impact upon the final state-to-state rate constants since it is well
recognized that even a small amount of quantum coherence between
states leads to a dramatic increase in the transition rate.
Fortunately, many of the papers presented in this proceedings address
these assumptions.  Approaches, such as the MCTDH method presented by
Thoss, the DFT based non-adiabatic molecular dynamics approach (NAMD)
developed by Prezhdo, for example, are important strides towards
achieving a molecular level understanding of complex photophysical
processes in light-emitting and light-harvesting materials.

\section{Acknowledgments}
This work was sponsored in part by the National Science Foundation and
by the Robert A. Welch Foundation.  The authors also thank the
organizers for putting together a highly stimulating conference in a
wonderful location.


%\printindex

\end{document}